\documentclass[12pt]{article}
\usepackage{latexsym}
\usepackage{graphics}
\usepackage{epsfig,amssymb,amsmath,graphicx,amsfonts}
\usepackage[dvips]{color}

\makeatletter
\newcommand{\singlespacing}{\let\CS=\@currsize\renewcommand{\baselinestretch}{1}\tiny\CS}
\newcommand{\oneandahalfspacing}{\let\CS=\@currsize\renewcommand{\baselinestretch}{1.25}\tiny\CS}
\newcommand{\doublespacing}{\let\CS=\@currsize\renewcommand{\baselinestretch}{1.5}\tiny\CS}

\newcommand{\expect}{{\Bbb {E}}}
\newcommand{\df}{\stackrel{{\rm def}}{=}}
\newtheorem{theorem}            {Theorem}
\newtheorem{lemma}              [theorem]{Lemma}

\begin{document}
\title{Two Proofs of the Fisher Information Inequality \\ via Data Processing Arguments}
\author{Tie~Liu and Pramod Viswanath}
\date{March 29, 2005}
\maketitle

\abstract{Two new proofs of the Fisher information inequality
(FII) using data processing inequalities for mutual information
and conditional variance are presented.}

\section{Introduction}
In parameter estimation problems, the Fisher information matrix of
a measurement $\mathbf{X}$ relative to a vector parameter
$\boldsymbol{\theta}$ is defined as
\begin{equation}
\mathbf{J}(\mathbf{X};\boldsymbol{\theta}) \df \mathrm{COV}\left\{
\frac{\partial}{\partial \boldsymbol{\theta}}\ln
f_{\boldsymbol{\theta}} (\mathbf{X})\right\} \label{eq:FI-G}
\end{equation}
where $\left\{f_{\boldsymbol{\theta}}(\mathbf{x})\right\}$ is a
family of probability density functions of $\mathbf{X}$
parameterized by $\boldsymbol{\theta}$, and
$\mathrm{COV}\{\cdot\}$ denotes the covariance matrix calculated
with respect to $f_{\boldsymbol{\theta}}(\mathbf{x})$. A special
form of the Fisher information matrix that shows up regularly in
information theory \cite{Joh} and physics \cite{Fri} is the Fisher
information matrix of a random vector with respect to a
translation parameter:
\begin{equation}
\mathbf{J}(\mathbf{X}) \df
\mathbf{J}(\boldsymbol{\theta}+\mathbf{X};\boldsymbol{\theta}) =
\mathrm{COV}\left\{\rho_{\mathbf{X}}(\mathbf{X})\right\}
\label{eq:FI-S} \end{equation} where the score function
$\rho_{\mathbf{X}}$ is defined as
\begin{equation}
\rho_{\mathbf{X}}(\mathbf{x}) \df \frac{\partial} {\partial
\mathbf{x}}\ln f(\mathbf{x}),
\end{equation}
and $f(\mathbf{x})$ is the probability density function of the
random vector $\mathbf{X}$. Unlike in the general definition
(\ref{eq:FI-G}), this special form of the Fisher information
matrix is a function of the probability density of the random
vector alone, and not of its parametrization.

Let $N_1$ and $N_2$ be two independent random variables with
probability density functions in the real line $\mathcal{R}$. The
classical Fisher information inequality (FII) states that
\begin{equation}
(a+b)^2 J(N_1+N_2) \leq a^2 J(N_1) + b^2 J(N_2), \quad \forall \;
a,\, b \geq 0. \label{eq:FII}
\end{equation}
Choosing $a=1/J(N_1)$ and $b=1/J(N_2)$, we obtain from
(\ref{eq:FII}) that
\begin{equation}
\frac{1}{J(N_1+N_2)} \geq \frac{1}{J(N_1)}+\frac{1}{J(N_2)},
\label{eq:FII-S}
\end{equation}
where the equality holds if and only if both $N_1$ and $N_2$ are
Gaussian. Compared with (\ref{eq:FII}), (\ref{eq:FII-S}) is
usually thought of as the canonical form of the classical FII.
\vspace{5pt}

The classical Stam-Blachman proof \cite{Sta59,Bla65} of the FII
relies on the following conditional-mean representation of the
score function for the sum of two independent random variables:
\begin{equation}
\rho_{N_1+N_2}(n)=\expect[\rho_{N_i}(N_i)|N_1+N_2=n], \quad i
=1,2, \label{eq:score}
\end{equation}
and then applies the Cauchy-Schwartz inequality. Although the
proof is direct and concise, it does not bring any
\emph{operational} meaning to the FII. \vspace{5pt}

In the excellent contribution \cite{Zam98}, Zamir showed that the
FII can be proved using the following data processing inequality
for Fisher information:
\begin{equation}
\mathbf{J}(\mathbf{Y};\boldsymbol{\theta}) \preceq
\mathbf{J}(\mathbf{X};\boldsymbol{\theta}) \label{eq:DPI-FI}
\end{equation}
if $\boldsymbol{\theta} \rightarrow \mathbf{X} \rightarrow \mathbf{Y}$
satisfies the chain relation
\begin{equation}
f(\mathbf{x},\mathbf{y}|\boldsymbol{\theta})
=f_{\boldsymbol{\theta}}(\mathbf{x})f(\mathbf{y}|\mathbf{x}),
\label{eq:chain}
\end{equation}
i.e., the conditional distribution of $\mathbf{Y}$ given
$\mathbf{X}$ is no longer a function of the parameter
$\boldsymbol{\theta}$. In \cite{Zam98}, Zamir considered the
parameter estimation model:
\begin{equation}
\left\{\begin{array}{lll}
X_1 & = & a \theta + N_1 \\
X_2 & = & b \theta + N_2
\end{array} \right.
\label{eq:model-Zamir}
\end{equation}
where $a$ and $b$ are two arbitrary nonnegative real numbers. Note
that $\theta \rightarrow (X_1,X_2) \rightarrow X_1+X_2$ satisfies
the chain relation (\ref{eq:chain}) in a trivial way. By the data
processing inequality (\ref{eq:DPI-FI}) for Fisher information, we
have
\begin{equation}
J(X_1+X_2;\theta) \leq J(X_1,X_2;\theta). \label{eq:Zamir}
\end{equation}
Thus, the desired inequality (\ref{eq:FII}) can be obtained by
substituting the parameter estimation model (\ref{eq:model-Zamir})
into (\ref{eq:Zamir}). Moreover, it can be seen that the
difference between the two sides of (\ref{eq:FII-S}) corresponds
to the loss in the Cram\'{e}r-Rao bound due to the ``processing"
in a certain linear additive noise model for parameter estimation.
\vspace{5pt}

It is worthy of mentioning that an identical argument without
assuming the independence between $N_1$ and $N_2$ proves a
generalization of the classical FII to the dependent-variable
case:
\begin{equation}
(a+b)^2J(N_1+N_2) \leq [a \;\; b] \mathbf{J}(N_1,N_2) [a \;\;
b]^t. \label{eq:FI-D}
\end{equation}
This result was initially proved in \cite[Th.~2]{Joh04} using the
conditional-mean representation of the score function for the sum
of two \emph{dependent} random variables.

\section{New Proofs of the FII}
Data processing is a general principle in information theory, in
that any quantity under the name ``information" should obey some
sort of data processing inequality. In this sense, Zamir's data
processing inequality for Fisher information merely pointed out
the fact that Fisher information bears the real meaning as an
information quantity. Interestingly enough, at the very beginning
of \cite{Zam98}, Zamir also pointed out that the data processing
principle applies to mutual information and conditional variance
as well. Specifically, if random variables $W \rightarrow X
\rightarrow Y$ form a Markov chain, the mutual information among
them satisfies
\begin{equation}
I(W;Y) \leq I(W;X), \label{eq:DPI-MI}
\end{equation}
and the conditional variances satisfy
\begin{equation}
\mathrm{VAR}[W|Y] \geq \mathrm{VAR}[W|X]
\label{eq:DPI-V}
\end{equation}
where $\mathrm{VAR}[W|X] \df \expect[(W-\expect[W|X])^2]$. The
main purpose of this note is to provide two new proofs of the FII
using the more familiar data processing inequalities
(\ref{eq:DPI-MI}) and (\ref{eq:DPI-V}), respectively.

\subsection{A Communications Proof}
Consider the communication model:
\begin{equation}
\left\{\begin{array}{lll}
X_1 & = & a\sqrt{t} W + N_1 \\
X_2 & = & b\sqrt{t} W + N_2
\end{array} \right., \quad t > 0
\end{equation}
where $W$ is standard Gaussian and $W$, $N_1$ and $N_2$ are
pairwise independent. By the scalar De-Bruijn identity
\cite{Bla65}, we have
\begin{eqnarray}
I(W;X_1) & = & \frac{a^2t}{2}J(N_1)+o(t), \label{eq:I(W;X_1)} \\
I(W;X_2) & = & \frac{b^2t}{2}J(N_2)+o(t), \label{eq:I(W;X_2)} \\
\mbox{and} \quad I(W;X_1+X_2) & = &
\frac{(a+b)^2t}{2}J(N_1+N_2)+o(t) \label{eq:I(W;X_1+X_2)}
\end{eqnarray}
where $\frac{o(t)}{t} \rightarrow 0$ in the limit as $t \downarrow
0$. Note that $W \rightarrow (X_1,X_2) \rightarrow X_1+X_2$ forms
a trivial Markov chain. By the data processing inequality
(\ref{eq:DPI-MI}) for mutual information, we have
\begin{eqnarray}
I(W;X_1+X_2) & \leq & I(W;X_1,X_2) \\
& = & I(W;X_1) + I(W;X_2|X_1) \\
& \leq & I(W;X_1) + I(W;X_2)
\label{eq:De-Bruijn}
\end{eqnarray}
where the last inequality follows from $I(W;X_2|X_1) \leq
I(W;X_2)$ because of the Markov chain $X_2 \rightarrow W
\rightarrow X_1$ \cite[p.~33]{CT}. Substituting
(\ref{eq:I(W;X_1)}), (\ref{eq:I(W;X_2)}) and
(\ref{eq:I(W;X_1+X_2)}) into (\ref{eq:De-Bruijn}), we obtain
\begin{equation}
(a+b)^2tJ(N_1+N_2) \leq a^2tJ(N_1)+b^2tJ(N_2)+o(t).
\end{equation}
Dividing both sides by $t$ and letting $t \downarrow 0$, we obtain
the desired inequality (\ref{eq:FII}). This completes the proof of
the classical FII using the data processing inequality for mutual
information. \vspace{5pt}

\noindent {\bf Remark.} The above proof can still go through even
without assuming $W$ is Gaussian. This is because, as mentioned in
the last paragraph of \cite[Sec.~III]{DCT91}, the scalar De-Bruijn
identity holds for any $W$ whose first four moments coincide with
those of Gaussian one.

\subsection{A Bayesian Estimation Proof}
Consider the Bayesian estimation model:
\begin{equation}
\left\{\begin{array}{lll}
X_1 & = & N_1 + \sqrt{at} W_1 \\
X_2 & = & N_2 + \sqrt{bt} W_2
\end{array} \right., \quad t > 0
\label{eq:Bayesian}
\end{equation}
where $W_1$ and $W_2$ are stand Gaussian random variables, and
$N_1$, $N_2$, $W_1$ and $W_2$ are pairwise independent. The
following lemma provides the needed connection between conditional
variance and Fisher information.

\begin{lemma} \label{lemma}
Let $N$ and $W$ be two independent random variables. Assuming $W$
is Gaussian with zero mean and variance $\sigma^2$, we have
\begin{equation}
J(N+W) = \frac{1}{\sigma^4}\left\{\sigma^2-\mathrm{VAR}[N|N+W]\right\}.
\end{equation}
\end{lemma}

\noindent {\bf Proof:} See Appendix \ref{app}. \hfill
$\blacksquare$ \vspace{10pt}

\noindent {\bf Remark.} Our proof uses the conditional-mean
representation of the score function for the sum of two
independent random variables, which suggests a natural connection
between conditional-mean estimators and Fisher information. We
mention here that Lemma \ref{lemma} may also be deduced from a
recent result of Guo, Shamai and Verd\'{u} \cite[Th.~1]{GSV05}, in
conjunction with the scalar De-Bruijn identity. Therefore, our
proof can be viewed as an alternative proof of the result of Guo,
Shamai and Verd\'{u}.

\vspace{5pt} Applying Lemma \ref{lemma} to model
(\ref{eq:Bayesian}), we obtain
\begin{eqnarray}
\mathrm{VAR}[N_1|X_1] & = & a t - a^2 t^2J(X_1), \label{eq:VAR[N_1|X_1]} \\
\mathrm{VAR}[N_2|X_2] & = & b t - b^2 t^2J(X_2), \label{eq:VAR[N_2|X_2]} \\
\mbox{and} \quad \mathrm{VAR}[N_1+N_2|X_1+X_2] & = & (a+b)t-(a+b)^2t^2J(X_1+X_2).
\label{eq:VAR[N_1+N_2|X_1+X_2]}
\end{eqnarray}
Note that $N_1+N_2 \rightarrow (X_1,X_2) \rightarrow X_1+X_2$
forms a trivial Markov chain. By the data processing inequality
(\ref{eq:DPI-V}) for conditional variance, we have
\begin{eqnarray}
\mathrm{VAR}[N_1+N_2|X_1+X_2] & \geq & \mathrm{VAR}[N_1+N_2|X_1,X_2] \\
& = & \mathrm{VAR}[N_1|X_1] + \mathrm{VAR}[N_2|X_2]
\label{eq:model-GSV05}
\end{eqnarray}
where the last equality follows from the fact that $N_1$, $N_2$,
$W_1$ and $W_2$ are pairwise independent. Substituting
(\ref{eq:VAR[N_1|X_1]}), (\ref{eq:VAR[N_2|X_2]}) and
(\ref{eq:VAR[N_1+N_2|X_1+X_2]}) into (\ref{eq:model-GSV05}), we
obtain
\begin{equation}
(a+b)^2J(X_1+X_2) \leq a^2J(X_1)+b^2J(X_2).
\label{eq:GSV05}
\end{equation}
Note that $J(X_1)$, $J(X_2)$ and $J(X_1+X_2)$ approach $J(N_1)$,
$J(N_2)$ and $J(N_1+N_2)$, respectively, in the limit as $t
\downarrow 0$. The desired inequality (\ref{eq:FII}) thus follows
from (\ref{eq:GSV05}) by letting $t \downarrow 0$. This completes
the proof of the classical FII using the data processing
inequality for conditional variance. \vspace{5pt}

\noindent \textbf{Remark.} As in Zamir's proof, the necessity of
the equality condition in (\ref{eq:FII-S}) does not follow easily
in either of the new proof. As a matter of fact, it becomes less
apparent due to the limiting argument used in both proofs.

\begin{appendix}
\section{Proof of Lemma \ref{lemma}}
\label{app} Let $X=N+W$. Since score functions always have zero
mean, the Fisher information of $X$ can be written as
\begin{equation}
J(X) = \expect[\rho_X^2(X)]. \label{eq:app1}
\end{equation}
By the conditional-mean representation of the score function for
the sum of two independent random variables, we have
\begin{eqnarray}
\rho_X(x) & = & \expect[\rho_W(W)|X=x] \\
& = & \frac{1}{\sigma^2}\expect\left[-W|X=x\right] \label{eq:app2}\\
& = & \frac{1}{\sigma^2}\left\{\expect[N|X=x]-x\right\}
\label{eq:app3}
\end{eqnarray}
where (\ref{eq:app2}) holds because $W$ is Gaussian with zero mean
and variance $\sigma^2$ so we have $\rho_W(w)=-w/\sigma^2$. It
then follows from (\ref{eq:app1}) and (\ref{eq:app3}) that
\begin{eqnarray}
J(X) & = &
\frac{1}{\sigma^4}\expect\left[\left(\expect[N|X]-X\right)^2\right]
\\
& = & \frac{1}{\sigma^4}\expect\left[\left(W+
(N-\expect[N|X])\right)^2\right]
\\
& = &
\frac{1}{\sigma^4}\left\{\sigma^2+\mathrm{VAR}[N|X]+2\expect[W(N-\expect[N|X])]\right\}
\label{eq:app4}.
\end{eqnarray}
Further note that
\begin{eqnarray}
\hspace{-20pt} \expect[W(N-\expect[N|X])] & = &
-\expect[(N-X)(N-\expect[N|X])]
\\
\hspace{-20pt} & = & -\expect[\left((N-\expect[N|X])+(\expect[N|X]-X)\right)(N-\expect[N|X])] \\
\hspace{-20pt}& = & -\mathrm{VAR}[N|X] \label{eq:app5}
\end{eqnarray}
because, by the orthogonality principle \cite{SW}, we have
$\expect[(\expect[N|X]-X)(N-\expect[N|X])]=0$. Substituting
(\ref{eq:app5}) into (\ref{eq:app4}), we obtain the desired
representation
\begin{equation}
J(X) =
\frac{1}{\sigma^4}\left\{\sigma^2-\mathrm{VAR}[N|X]\right\}.
\end{equation}
This completes the proof of Lemma \ref{lemma}.
\end{appendix}

\end{document}